\documentclass[12pt]{article}
\usepackage{a4wide}
\usepackage{amsmath,amssymb,bbm}

\def\alphap{{\alpha'}} 
\def\Zbf{{\bf Z}} 
\def\coeff#1#2{\frac{#1}{#2}}
\def\half{\frac12}    
\def\ie{{\it i.e.}}
\newcommand{\abs}[1]{\lvert #1\rvert}
\newcommand{\zb}{\bar{z}}
\newcommand{\ab}{\bar{a}}

\begin{document}
\setcounter{page}{1}
%


%

\def\pct#1{(see Fig. #1.)}

\begin{titlepage}
\hbox{\hskip 12cm NIKHEF/2003-014  \hfil}
\hbox{\hskip 12cm IMAFF-FM-03/17  \hfil}
\hbox{\hskip 12cm hep-th/0310295 \hfil}
\vskip 1.4cm
\begin{center}  {\Large  \bf  On Orientifolds of c=1 Orbifolds}

\vspace{1.8cm}
 
{\large \large T.P.T. Dijkstra, B. Gato-Rivera\footnote{Permanent address: Instituto de Matem\'aticas y F\'\i sica
Fundamental, CSIC, Serrano 123, Madrid 28006 Spain.},  F. Riccioni\footnote{New address: 
DAMTP, Wilberforce Road, Cambridge CB3 0WA, UK.} and A.N. Schellekens}
\vspace{0.6cm}

{\sl NIKHEF \\
P.O. Box 41882 \ \ 1009 DB \ \ Amsterdam \\
The Netherlands}
\end{center}
\vskip 1.5cm

\abstract{The aim of this paper is to study orientifolds of $c=1$ conformal field theories.
A systematic analysis of the allowed orientifold projections for $c=1$ orbifold conformal
field theories is given. We compare the Klein bottle amplitudes obtained at rational points
with the orientifold projections that we claim to be consistent for any value of the orbifold 
radius. We show that the recently obtained Klein bottle amplitudes corresponding to exceptional 
modular invariants, describing bosonic string theories at fractional square radius, are
also in agreement with those orientifold projections.
}

\vfill
\end{titlepage}
\makeatletter
\@addtoreset{equation}{section}
\makeatother
\renewcommand{\theequation}{\thesection.\arabic{equation}}
\addtolength{\baselineskip}{0.3\baselineskip} 

\section{Introduction}\label{sec:introduction}
The c=1 orbifold CFTs on closed, oriented Riemann surfaces have been studied extensively
\cite{DVV}\cite{DVVV}\cite{Ginsparg}\cite{Kiritsis} because they provide
simple but non-trivial examples of various features of conformal field theory.
It has long been believed that all c=1 bosonic
theories in the closed oriented case were known:
they either belonged to the famous continuous moduli space of the
circle and its $Z_2$ orbifold, or one of three discrete points discovered in \cite{Ginsparg}.
The completeness proof \cite{Kiritsis} was based on certain assumptions, and more recently
counterexamples have been conjectured to exist \cite{RuWa}. The open, unoriented
case has received considerably less attention.
Early results can be found in \cite{ps}, reviewed
recently in \cite{AngelSagn}, and there are some remarks on such theories in
the appendix of \cite{cda}.
Our purpose here is a systematic study of the set of theories
that is obtained if one allows open and unoriented surfaces. We will study this problem
for continuous values of the radius $R$ and at rational points, and match the results.

For the circle theories the solution to this problem is known. On oriented, closed
surfaces the moduli space is a line parametrized by the radius $R$ of the circle, with
a T-duality identification of $R$ and $\alpha'/R$. On unoriented surfaces this line
splits into two lines. In one of the two T-dual pictures the two lines are characterized
by having $\rm{O}0$ planes of either the same tension or opposite tension on opposite points
of the circle, in the other the circle is covered by $\rm{O}1$ planes.
This matches precisely and unambiguously with the results from
rational CFT, as we will see in slightly more detail below.

Studying this problem on the orbifold line is interesting for a variety of reasons: the
set of modular invariant partition functions (MIPFs) 
is richer, there is an interesting one-to-one
map between the rational c=1 orbifolds at radius $R^2=\alpha' N$ and the $D_N$ WZW-models at
level 2 \cite{Cloning} and  the orbifolds have a twist field degeneracy \cite{DVVV} which is
resolved only in the rational points. Finally we are interested
in understanding 
the puzzling results of \cite{NunoSch}, showing that in certain rational points the orbifold CFT
does not admit a canonical Klein bottle projection (with all coefficients equal to 1).  There
are non-canonical solutions, but it is not immediately clear how they would fit into a continuum description.

Two types of approaches to boundary RCFT are used in this paper.
In \cite{FHSSW} a general formula was given for all reflection coefficients
and crosscap coefficients for all simple current modular invariants \cite{KrS} and a large class of
orientifold projections. This formula is consistent (at least in the
sense of yielding integral coefficients, see \cite{FHSSW} and \cite{Lenn}; furthermore
consistency for all resulting oriented amplitudes was demonstrated recently in \cite{FuchsDZ}), but there is no proof that it is
complete. In \cite{SchStan} a set of polynomial equations was written down which can be solved
for annulus, Moebius and Klein bottle coefficients. This method is complete, but it
is not known if all solutions are physical.
Our goal is to do
a systematic analysis of all known orbifold RCFT's and compare them with the continuum,
to check if the method of \cite{FHSSW} misses anything, and if the method of \cite{SchStan} produces
anything manifestly unphysical. The orbifolds are an ideal laboratory for this because the continuum can
be obtained
explicitly, and RCFT's are known in all rational points. The circle has all
these features as well, but lacks some non-trivial structure related to the fixed points, as well
as exceptional modular invariants.  

A prerequisite for this work is a complete description of orientifolds in the continuous case. 
Although there is a large body of work of orientifolds of tori and orbifolds, as far as we know
the complete and explicit answer for the $c=1$ orbifolds is not available yet, although a partial result
can be found in \cite{AngelSagn}.
We find a total of four different
allowed Klein bottle amplitudes and corresponding orientifold projections.


The structure of the paper is the following. In Section 2 we review the results concerning
the bosonic string on a circle and the orientifolds thereof. In Section \ref{sec:orbifolds}
we consider the
orbifold case. First, in subsection 3.1, we study orientifolds for rational radius.
A parametrization of the allowed Klein bottles in terms of RCFT characters leads to twelve
different choices. Then we enumerate the modular invariant partition functions (MIPFs), by
systematically combining all known types and checking closure under multiplication.
For each MIPF we perform a systematic construction of
U-NIMreps \cite{OriMat} (U-NIMreps are sets of Annulus, M\"obius and Klein bottle amplitudes that
satisfy the aforementioned polynomial equations; they are NIMreps
\cite{DiF}\cite{PSSC}\cite{BPPZ} extended with data concerning unoriented surfaces.) This method
is limited in practice to a finite number of primaries, but we can extend it far enough to uncover
the complete picture, and furthermore we supplement it with the formula of \cite{FHSSW}, which has
no such limitation. We conclude that for integer values of $R^2/\alpha' $ six different cases can
be distinguished, corresponding to four distinct continuum Klein bottle amplitudes. As a result
of these computations we obtain a set of boundary and crosscap coefficients which we use in
subsection 3.1.4 to study the localization of branes and O-planes, as well as the Chan-Paton groups
in the various cases. In subsection 3.2 we study the case of orientifolds at continuous
radius from a geometric point of view, and confirm that the four Klein bottle amplitudes
are indeed the only possible ones. In subsection 3.3 we consider the case of certain non-integer
rational radii (exceptional MIPFs) and show that, despite their unexpected features,
the results of \cite{NunoSch} are precisely in agreement with the four
orientifold projections. In Section 4 we summarize our conclusions.
In the appendix we discuss a variety of orbifold maps
needed to obtain distinct orbifold theories that exist in the rational case.

\section{Summary of circle results}\label{sec:summ-circle-results}

In this section we want to review know results about orientifolds of the bosonic string 
on a circle, in order to set up the discussion for the orbifold case and to introduce 
some notations. Orientifolds of circle compactifications for irrational values of the radius 
appeared for the first time in \cite{bps}, while additional references are 
\cite{dp}\cite{abpss}\cite{w}\cite{FuchsFu}\cite{bgs}\cite{Marijn} (see \cite{AngelSagn} 
for a review).
The partition function for a bosonic closed oriented string compactified on
a circle is
\begin{equation}\label{part_circle}
Z(R)=\frac{1}{\eta\bar\eta} \sum_{m,n}
q^{\frac{\alphap}{4}\left(m/R+nR/\alphap\right)^2} 
\bar q^{\frac{\alphap}{4}\left(m/R-nR/\alphap\right)^2},
\end{equation}
where we denote with $n$ the winding and $m/R$ the Kaluza-Klein (KK) momentum along the
circle.
A well known feature of this partition function is the fact that it is
invariant under the exchange $R \leftrightarrow \alphap/R$. From the point of
view of the extended CFT that describes the above theory, 
the partition function $Z(R)$ hides chiral
information. For each value of $R$ there are actually two theories with the same
partition function, which are each others T-dual, one with momentum states
$|\sqrt{\alphap\over2}(m/R+nR/\alphap),\sqrt{\alphap\over2}(m/R-nR/\alphap)
\rangle$, and one with states
$|\sqrt{\alphap\over2}(m/R+nR/\alphap),\sqrt{\alphap\over2}(-m/R+nR/\alphap)
\rangle$. The first of these is obtained in a genuine compactification on a
circle of radius $R$.

The allowed orientifold projections must respect the operator product
expansion of the CFT (we will only consider orientifold projections of
order 2).  This implies in particular that the operator $\partial X \bar
\partial X$ must be transformed into itself, and that the vertex operators
corresponding to the momentum and winding states must transform with a factor
$(\epsilon_1)^m (\epsilon_2)^n$, with $\epsilon_1$ and $\epsilon_2$ equal
to $\pm 1$.  Hence one can associate four consistent Klein bottles with
this partition function, namely\footnote{We assume here that the relation between
the orientifold projection and the Klein bottle is straightforward. This implies
in particular that the Klein bottle coefficients are preserved under fusion, {\it i.e}\ that
the ``Klein bottle constraint" is satisfied. There are examples where this constraint
is not satisfied (see {\it e.g}\ \cite{HSSA}) and which require further thought, but
this problem does not occur for any of the c=1 U-NIMreps.}

\begin{subequations}
\begin{gather}
\frac{1}{\eta(2i\tau_2)} \sum_m q^{\frac{\alphap}{2} (m/R)^2} \equiv 2 K_{+00}(R)\\
\frac{1}{\eta(2i\tau_2)} \sum_m (-1)^m q^{\frac{\alphap}{2} (m/R)^2}  \equiv 2 K_{-00}(R) \\
\frac{1}{\eta(2i\tau_2)} \sum_n q^{\frac{1}{2\alphap} (nR)^2}  \equiv 2 K_{0+0}(R)  \\
\frac{1}{\eta(2i\tau_2)} \sum_n (-1)^n q^{\frac{1}{2\alphap} (nR)^2} \equiv 2
K_{0-0}(R) \ .  
\end{gather}
\end{subequations}
The functions $K$ will be introduced later. The Klein bottle amplitude is
subject to the constraint that $\half(Z(R)+K(R))$ expands into 
non-negative integers.
The first two Klein bottles can be combined with the diagonal theory, the
other two with its T-dual.
The positions of the orientifold planes can be derived from the transverse channel Klein
bottle amplitudes by dimensional analysis. In the first two cases the transverse amplitude
describes a closed string propagating between two O1 planes. There are two different 
O1 planes. In the first case, the resonance term permits only to even winding states
to propagate in the transverse channel, while in the second case only odd winding states 
contribute, and thus the configuration has vanishing tension, since
the graviton does not propagate in the transverse channel.
We call these two configurations $\rm{O}1_+ \oplus \rm{O}1_+$ and $\rm{O}1_+ \oplus \rm{O}1_-$ respectively.
In the first case, the orientifold projection maps $X$ to itself, while in the second case it maps $X$ 
to $X + \pi R$. Both these maps square to the identity because of the periodicity of the circle, 
and have no fixed points. 
The other two cases are the T duals of the former.
After T duality, the orientifold projection
acts as a $Z_2$ orbifold on the circle coordinate, so that the model lives on a segment, with O0 planes at
the endpoints. In the third case, the orientifold projection maps $X$ to $-X$, while in the fourth case it maps
$X$ to $-X +\pi R$. Both these projections have fixed points, where the O0-planes are located.
More precisely, the third case corresponds to the configuration
$\rm{O}0_+ \oplus \rm{O}0_+$, in which the two O-planes have the same tension, while the 
fourth case corresponds to  $\rm{O}0_+ \oplus \rm{O}0_-$, in which the O-planes have opposite tension.

The corresponding rational CFTs are obtained by setting $R^2=\alphap N$. All primaries of this
CFT are simple currents, forming a $Z_{2N}$ discrete group. We denote the generator of $Z_{2N}$ as $J$.
All MIPFs
are simple current invariants, and they are in one-to-one correspondence with
the subgroups of $Z_{2N}$ generated by even powers of $J$, which in their turn are
in one-to-one correspondence with the divisors of $N$.
If $m$ is
a divisor of $N$, the MIPF belonging to the subgroup
generated by $J^{2m}$ corresponds to a circle with radius $R^2=\alphap N/m^2$.  
We will refer to this partition functions as $Z_{\rm circle}(m,N)$, with the convention
that $Z_{\rm circle}(N,N)$ is the charge conjugation invariant and $Z_{\rm circle}(1,N)$
the diagonal one.
For every rational radius
$R^2=\alphap p/q$ (with $p$ and $q$ relative prime)
there is an infinite number of rational CFTs describing it, namely $Z(pr,pqr^2)$,
for any $r \in \Zbf$. For $r > 1$ the
corresponding partition functions involve extensions of the chiral algebra. 

For all these MIPFs the allowed crosscap and boundary coefficients follow from
the general formula presented in \cite{FHSSW} (summarizing and extending earlier work
in \cite{pre-FHSSW}), which in the special case $Z_{N,N}$ reduces
to the well-known boundary state of Cardy \cite{cardy} combined with the crosscap state due
to the Rome group \cite{PSS_Gamma-P}.

By Fourier analyzing the closed string scattering amplitudes from the
boundary and crosscap states
(a procedure that was pioneered in \cite{Vecchia}, and applied in  \cite{FFFS} 
to boundary states of WZW models and in
 \cite{HSS} to crosscap states (see also \cite{Brunner}\cite{Bachas}))
one can localize the D-branes and O-planes on the circle. To do this one
multiplies the boundary and crosscap amplitude with a factor $e^{ikx/R}$ and sums over
all values of $k$ in the primary range $-N \leq k < N$. The resulting function of $x$
has peaks that get more pronounced with increasing $N$, and are interpreted as the brane
and plane positions.
>From the continuous
point of view, we expect D1 branes for the diagonal invariant (corresponding
to a genuine circle compactification, with the space-filling brane wrapped
around the circle), which turn into D0 branes for the charge
conjugation invariant, the T-dual of the former.

In the rational CFT one finds the following.
The MIPF $Z(m,N)$ admits $2m$ boundaries, each of
which is localized at $n=N/m$ distinct points on the circle.    To make sense
of this we introduce the dual radius $\hat R=\alphap/R$, which is the
relevant quantity because D0 branes live
on the dual circle.
The $n$-fold multiplicity is an indication of the fact that the original circle
of dual radius $\hat R$ is to be interpreted as  an $n$-fold cover of a circle of radius
$\hat R/n$.
We may label the
boundaries by integers $a=0,\ldots,m-1$, such that they are localized at points
$\left({a+2m \ell \over 2N}\right) 2\pi \hat R$, $\ell=0,\ldots,n-1$.  For $m=N, n=1$
(the charge conjugation invariant) this yields $2N$
D0 branes equally distributed over the circle; for $m=1, n=N$
(the diagonal invariant) this gives two
branes localized simultaneously on $N$ equally distributed points, one brane
on the odd points and one on the even points. This is the RCFT realization of a D1 brane.

In the continuum the D0
can be localized anywhere on the circle, whereas in the rational CFT
description their positions are quantized. One can approach the continuum
results either by using deformations of the boundary CFT \cite{ReckScho}, or by allowing
boundaries that break some of the extended symmetries that characterize
the rational CFT. This can be done by obtaining the circle at some dual radius $\hat R$
from a circle at dual radius $\hat Rr$ by extending the CFT of the latter. The RCFT notion
of ``completeness of boundaries" \cite{PSS}, when applied to MIPFs of
extension type, automatically implies the presence of boundaries that break the
extended symmetries. Indeed, if we consider $Z(pr,pqr^2)$ we find
$2pr$ distinct boundaries, each localized in $qr$ points on the circle of
radius $\hat Rr$. This circle is an $r$-fold cover of the circle of radius $\hat R$, so that
on the latter circle we now have $2pr$ distinct branes each localized in
$q$ points. Of these, $2p$  coincide with the ones found for $r=1$,
and the remaining ones occupy intermediate positions. In the limit $r\to \infty$,
$\hat R$ fixed we approach the continuum result.

Similar results hold for crosscaps.
The formalism of \cite{FHSSW} allows two ways of modifying the crosscap
coefficient
for a given MIPF. The formula for the crosscap coefficient is, up to normalization
\begin{equation}
\Gamma_i \propto \sum_{L \in G} \eta(K,L) P_{KL,i} \ ,
\end{equation}
where $i$ is the Ishibashi label, $P$ the $P$-matrix ($P=\sqrt{T}ST^2S\sqrt{T}$).
Here $K$ is a simple current (subject to a condition given below),
$\eta(K,L)$ a set of signs satisfying the constraint
\begin{equation}
\eta(k, L) = e^{\pi i (h_K-h_{KL})} \beta_K(L) \quad ,
\end{equation}  
where $\beta_K(L)$ is a set of phases solving the relation
\begin{equation}
\beta_K(LJ) = \beta_K(L)\beta_K(J) e^{-2\pi i X(L,J)}, 
\end{equation}    
where $X$ is the rational bihomorphism that specifies the MIPF, as defined in \cite{KrS}.
This relation does not fix the phases completely:
for every independent even cyclic factor of the simple current group $G$, there is a free sign.
These free signs are called the ``crosscap signs".
The current $K$ (for historical reasons called the ``Klein bottle current")
must be local w.r.t. the currents of order two in $G$,
and currents that differ by elements of $G$ or by squares of simple currents yield
equivalent crosscaps. 
These Klein bottle currents form, together with the crosscap signs, the set of allowed
crosscap modifications.

In the present case, it is not hard to see
that for each choice of $G$ there are just two solutions. If $G$ has even order, there is
a single crosscap sign choice, but there are no Klein bottle currents local w.r.t. $G$. If $G$ has
odd order $n$ there is no crosscap sign choice, but then the current $J^{n}$ is local
w.r.t. $G$ and is a non-trivial Klein bottle current. These two choices correspond
precisely to the two orientifold choices in the continuous case. In all cases one
of the orientifold choices leads to Klein bottle coefficients that are equal to +1
for all fields appearing diagonally.

The crosscap positions can be worked out in the same way as for D-branes. For the MIPF
$Z(m,N)$ we find that a crosscap state occupies $2n$ positions, twice as many as a boundary
state. These positions are $n$-fold identified. In the simplest case, $m=N, n=1$
there are two positions, diametrically opposite on the circle. In this case,
the crosscaps are characterized by the choice of Klein
bottle current $K=J^{k}$. Each $k$ corresponds to two $\rm{O}0$ planes localized at
$r={k\over 4N} 2\pi R$ and $r=({k+2N\over 4N}) 2\pi R$. If $k+N$ is even these two $\rm{O}0$
planes have the same tension, if $k+N$ is odd they have opposite tension. If $k$ is even,
the O-plane locations coincide with a brane position; if $k$ is odd the O-planes lie
between two brane positions.
Configurations
where $k$ differs by an even integer can be obtained from each other by rotating the circle,
in agreement with the fact that the corresponding Klein bottle currents are equivalent.
The T-dual configuration corresponds to the diagonal invariant, obtained by using the
simple current $J^2$. This MIPF admits just two Ishibashi states, 
and hence only two non-vanishing crosscap coefficients.
This is not sufficient to contain any information about localization, in agreement with the fact
that we expect the O-planes to be $\rm{O}1$ planes wrapping the circle (and analogously
for boundaries).

\section{Orbifolds}\label{sec:orbifolds}

We are considering the $c=1$ case, that is the real line modded out by
the group $G$ of reflections and translations, resulting in the segment $\mathbbm{R} /G = S^1 /\mathbbm{Z}_2$. Following
\cite{int-on-orb} we will denote the action of elements of this group on the string
coordinate as
\begin{equation} 
(\theta,n)\in G \ , \ \  (\theta,n) X = \theta X +  2 \pi n R \ , \qquad n \in \mathbb{Z} \ , \; \theta = \pm \ .
\end{equation}
Strings on an orbifold are closed if they are periodically identified {\em up to} an
element of this group:
\begin{equation}
X(\sigma+2 \pi) =  (\theta,n) X(\sigma) \ .
\end{equation}
$X$ is then twisted by the element $(\theta,n)$;  
this defines various sectors with different periodicity conditions on $X$.  Not all elements of $G$ give
rise to a different sector, a sector twisted by $g$
being the same as the one twisted by $h\,g\,h^{-1}$. 
Thus we get a sector for each conjugacy class.  One has the following
conjugacy classes: $(+,\pm \abs{m})$, $(-,\hbox{even})$ and
$(-,\hbox{odd})$. The first gives the circle periodicity conditions
with winding number $m$, where now the winding direction is no longer
significant. The last two cases give the twisted sectors. Note that in
these sectors the notion of winding is limited to being 'even' or
'odd'.

In the untwisted sector, $X$ has the same mode expansion as in the
circle, with the difference that now only the states that are
invariant under the map $X\to-X$ are present.  Denoting with $r$ the
operator that performs this map on the Hilbert space, the resulting
spectrum is obtained by applying the projector $\frac{1+r}{2}$ on the
circle states.

In the twisted sectors the mode expansion for $X$ is 
\begin{equation}
X = x_{o} +  \sum_{n} \frac{1}{n-\tfrac{1}{2}} \left(
  a_{n-\tfrac{1}{2}} z^{-n+\tfrac{1}{2}} + \ab_{n-\tfrac{1}{2}}
  \zb^{-n+\tfrac{1}{2}} \right) \ , \label{X_twisted}
\end{equation}
where $x_{o} = 0$ for the $(-,\hbox{even})$ sector, $x_{o} = \pi R$
for the $(-,\hbox{odd})$ sector.  The twisted sectors correspond thus to
states localized at the fixed points of the orbifold.  Also the states
created by the modes in \eqref{X_twisted} must be projected by the
operator $\frac{1+r}{2}$.

The resulting partition function is 
\begin{equation}\begin{split}\label{part_orb}
Z_{\rm orb}(R) 
&= \half Z_{\rm circle}(R)
+ \left| \frac{\eta}{\theta_2} \right|
+ \left| \frac{\eta}{\theta_3} \right| 
+ \left| \frac{\eta}{\theta_4} \right| \ .
\end{split}
\end{equation}
The first two are contributions of the untwisted sectors, 
while the last two are contributions of the twisted sectors.

\subsection{Orientifolds for rational radius}\label{sec:cft-approach}  
\subsubsection{Parametrizations of the Klein bottle}

In the CFT description, the allowed orientifold projections are limited
by the requirement of preservation of the OPE. Of most interest are the
projection signs of states appearing diagonally, since those signs affect
the Klein bottle. Again we find that $\partial X \bar \partial X$ must
transform into itself. This implies that ${\eta\over\theta_2}$ cannot
appear in the Klein bottle expression, since this
function represents the {\it difference} of contributions of the identity operator
and the operator $\partial X \bar \partial X$.
The OPE of $\partial X \bar
\partial X$ with a 
lowest weight twist field $\sigma(z,\bar z)$ yields the excited twist field $\tau(z,\bar z)$. Since
$\partial X \bar \partial X$ has projection sign $+$, the twist fields
$\sigma$ and $\tau$ must be projected in the same way. This implies the
absence of ${\eta\over\theta_3}$, which corresponds to the difference of
the two twist field labels.

An important issue is twist field degeneracy. The $c=1$ orbifold has two
twist fields (stemming from the fact we have two different twisted
sectors), denoted $\sigma_1$ and $\sigma_2$ (with weight
$(\coeff1{16},\coeff1{16})$) and two excited twist fields, $\tau_1$ and
$\tau_2$, with weight $(\coeff9{16},\coeff9{16})$. The labels 1 and 2 are
not distinguished by the Virasoro algebra. On any point on the orbifold
line the Virasoro algebra is extended by operators that are even polynomial
in $\partial X$ and its derivatives, the first one at weight 4 \cite{DVVV}.
But these operators do not distinguish the labels either, since $\partial
X$ itself does not. Only in the rational points there are operators that
distinguish the twist fields, namely the operators ${\rm
  cos}(\sqrt{2\over\alphap} m R X)$ that extend the chiral algebra to make
the CFT rational.  Hence we should regard these as states with multiplicity
2. The allowed Klein bottle coefficients in this sector are the $2, 0$ or
$-2$. The value 0 is allowed if the twist fields appear off-diagonally in
the partition function, or if they appear diagonally, but have opposite
Klein bottle projections.  Based on this information we arrive at the
following twelve choices for the Klein bottle
\begin{equation}\label{klein_orb}
K_{\epsilon_1\epsilon_2\epsilon_3}=
\half \frac{1}{\eta(2i\tau_2)} \sum_{k\in\Zbf} (\epsilon_1)^k q^{\frac{\alphap}{2}\left(\frac{k}{R}\right)^2}
+ \half \frac{1}{\eta(2i\tau_2)} \sum_{m\in\Zbf} (\epsilon_2)^m
q^{\frac{1}{2\alphap}\left(mR\right)^2}+ 
2\epsilon_3 \sqrt{ \frac{\eta}{\theta_4} } \ ,
\end{equation}
with $\epsilon_1=\pm1; \epsilon_2=\pm 1$ and $\epsilon_3=0,\pm 1$.  The same parametrization
can be used for the circle theory, provided one allows the value 0 for $\epsilon_1$ and
$\epsilon_2$.

As was the case for the circle, the allowable Klein bottles depend on the interpretation 
of the partition function. But in contrast to the circle theory, this interpretation is
discontinuous in $R$. This is due to the fact that we can distinguish the twist fields only
for rational $R$. In addition, for rational $R$ two orbifold fields appear that do
not exist for irrational values of $R$, namely the fields we denote as $\phi_1$ and $\phi_2$ and
that have conformal weight $\frac14 N$. 

\subsubsection{Enumeration of modular invariants}\label{Emi}

We will now study the orbifold at rational points in order to reduce the set of orientifold
projections. In the rational points and for a sufficiently small set of primaries we have
an additional tool at our disposal, namely the systematic search for NIMreps and U-NIMreps.

The orbifold of the circle of radius $R^2=\alphap N$ (or its T-dual) is the well-known
orbifold rational CFT with $N+7$ primaries. It has four simple currents, $1$, $\phi_1$, $\phi_2$
and the spin-1 current $\partial X$, forming a discrete group $\Zbf_4$ (for $N$ odd) 
or $\Zbf_2 \times \Zbf_2$  (for $N$ even). The remaining fields will be denoted
$\varphi^k$, $k=1,\ldots,N_1$, $\sigma_i, \tau_i (i=1,2)$, following \cite{DVVV}.
A lot is known about the MIPFs of these CFTs, but the result are scattered
throughout the literature, and for that reason we will give here an
enumeration of what is known.

In \cite{DVVV} it was observed that the
theory at radius $R^2=\alphap p/q$ or
$R^2=\alphap q/p$ has the same chiral algebra as the one at $R^2=\alphap pq $.
Hence it is described by a non-trivial MIPF of the theory at
$R^2=\alphap pq $.
This MIPF
is of exceptional type, except when $q$ and/or $p$ is equal to 2, in which case the
invariant is of simple current type. Just as
in the circle case, one can generalize this
by allowing $p$ and $q$ to have common factors.  In this way one
can obtain an infinite number of rational CFT realizations at any rational point on
the orbifold line.
Any of these MIPFs can be extended by the
simple current $\partial X$ to re-obtain the circle partition function.

But there are still more rational partition functions for every rational point, a fact
that can most easily be appreciated by using the fact that the
modular group representation of the orbifold CFT for $R^2=\alphap N$
is in
one-to-one correspondence with the $D_N$ WZW model at level 2 \cite{Cloning}.
In particular there is a one-to-one relation for partition functions, NIMreps and
U-NIMreps. While the aforementioned MIPFs describing orbifolds at
non-integer radii do not seem to have a
{\it raison d'etre} for the WZW-models, they do exist for these models as well.
The ones of automorphism type were discovered using
Galois symmetry in \cite{Galois}
(subsequently the WZW automorphisms were fully classified in \cite{GRW}); the extensions
were described in \cite{Cloning} using the aforementioned correspondence.
On the other hand, the $D_N$ WZW models (at any level) have
MIPFs related to Dynkin diagram automorphisms that imply the existence
of related invariants for the orbifolds. These are first of all the conjugation invariants,
which have an off-diagonal pairing of $\phi_1,\phi_2$, $\sigma_1,\sigma_2$ and
$\tau_1,\tau_2$. For odd $N$ this is the charge conjugation
invariant, which can also be described as a simple current invariant generated by $\phi_1$
(or, equivalently, $\phi_2$). For even $N$ this is an exceptional invariant
(for even $N$, the
simple current invariant
generated by $\phi_1$ gives an orbifold at reduced radius $N/4$).
If the MIPF involves an extension by $\phi_1$ or $\phi_2$ (which happens if
$p$ and $q$ are both even), there are still more possibilities, because
one can then conjugate the left and right chiral algebras independently.
As a result one obtains two symmetric and two asymmetric (heterotic) MIPFs.
Finally, for $N=4$ there are even
more MIPFs related to triality of $D_4$; there are 16 MIPFs in total.

For given $N \not=4$ the number of known modular invariants, obtained by combining all
the above, can be described as follows.
Let $p$ be a divisor of $N$ in the range $1 \leq p \leq \sqrt{N}$, and define $q=N/p$. The
number of known invariants is equal to $\sum_p M(p)$ where the sum is over all
divisors $p$ in this range, and $M(p)=3$ if either $p$ or $q$ is odd, $M(p)=5$ otherwise.
The multiplicity
three corresponds to the diagonal invariant, the conjugation invariant and the circle extension,
which we denote respectively as $Z_D(p,N)$, $Z_C(p,N)$ and $Z_X(p,N)$.
The multiplicity five corresponds to the four cases described above plus the circle
extension, denoted respectively as
$Z_{11}(p,N), Z_{22}(p,N), Z_{12}(p,N), Z_{21}(p,N)$ and $Z_X(p,N)$
(the circle extension includes both $\phi_1$
and $\phi_2$).       

Since the standard orbifold map $X \to -X$ yields just one orbifold theory for each T-dual
pair of circle theories, one has to consider more general orbifold maps to get the various
types of orbifold partition functions. These maps are discussed in the appendix.

\subsubsection{U-NIMreps}\label{subsec:unimreps}

Consider first $Z_C(1,N)$ and $Z_D(1,N)$. In all but one case
these MIPFs are C-diagonal or of simple current type, and
a set of Klein bottle coefficients can be
obtained from various previous papers. To deal with the remaining exceptional
 MIPF
($Z_C(1,N)$, $N$ even) , and
as a check on all the others, we solved the U-NIMrep polynomial
equations (for $N \leq 16$) to get the complete answer. This results
in the following six cases:

\begin{enumerate}
\item The diagonal invariant, with standard Klein bottle ($K_i=1$ for all
$i$).  For $N$ even, this is the standard Cardy-Rome case. For $N$ odd, it
is a simple current automorphism, as treated in \cite{FHSSW}, with suitable
choice of the crosscap sign. The resulting Klein bottle amplitude is
$K_{+++} $. This result was first obtained in \cite{ps} (see also
\cite{AngelSagn}).

\item The diagonal invariant, with Klein bottle coefficients $-1$
in the twisted sector.
For $N$ even, this is the same as the previous case but with
Klein bottle current 2.  For $N$ odd, it is the same as the
previous case, but with the opposite crosscap sign.
The resulting Klein bottle is $K_{++-}$.

\item The diagonal invariant, with Klein bottle currents $\phi_1$ or $\phi_2$.
This case exists only for $N$ even.  For $N$ odd, the diagonal invariant is
generated as a simple current automorphism of current $\phi_1$, and the
only allowed sign changes are the crosscap signs, which we already saw
above. For even $N$ one finds that all odd charged fields $\varphi^k$ get a
negative Klein bottle, \ie\ $K_{\varphi^k}=(-1)^k$. Furthermore either
$\sigma_1$ and $\tau_1$ or $\sigma_2$ and $\tau_2$ change sign, so that the
total twisted sector contribution cancels. The result is $ K_{-+0} $ (for
$N$ even only).

\item The charge conjugation invariant with standard Klein bottle (\ie\ all
$K_i = +1$ if $i$ appears diagonally).  The charge conjugation invariant
only differs from the diagonal one for $N$ odd. The effect is that
$\phi_1$, $\phi_2$ appear off-diagonally, and the same for the twisted
sector. The latter implies $\epsilon_3=0$. The absence of $\phi_1$,
$\phi_2$ in the Klein bottle amplitude implies $\epsilon_2=-1$, so that the
contribution of $\phi_i$ cancels between the first two terms. Hence we get
$ K_{+-0} $ (for $N$ odd only).

\item The charge conjugation invariant with non-standard Klein bottle (case 4
with simple current Klein bottle current $\phi_1$, which is equivalent to $\phi_2$).
This gives a sign flip for all odd charge fields, implying $\epsilon_1=-1$
This can be taken into account by inserting a $(-1)^k$ into the first sum. 
Since $N$ is odd, the $\phi_i$ contribution cancels between the first two terms
if and only if $\epsilon_2=+1$. Hence we get
 $  K_{-+0}$ (for $N$ odd only).

\item The conjugation invariant for even $N$. This is an exceptional invariant that
pairs $\phi_1$ with $\phi_2$, $\sigma_1$ with $\sigma_2$ and $\tau_1$ with $\tau_2$.
Here \cite{FHSSW} does not apply, but by solving the NIMrep conditions explicitly we find
only one NIMrep with one U-NIMrep. The Klein bottle has all allowed coefficients equal to 1. 
This yields $ K_{+-0}$ (for $N$ even only).
\end{enumerate}

We summarize these results in the following table. In the first column `D' denotes the
diagonal invariant, `C' the charge conjugation invariant, and `T' the twist field
conjugation invariant, in which $\phi_{i},\sigma_{i}$ and $\tau_i$ are off-diagonal.
In the fifth column we indicate the Chan-Paton group for the dominant branes ({\it i.e.}
the ones that are most numerous for large $N$). This will be explained in the
next subsection. The last column refers to the six cases listed above.

\vskip .3truecm
\begin{center}
~~~~~~~~~~~~~~~~\begin{tabular}{|l|c|c|c|c|c|} \hline 
Invariant &  $N$ & Boundary/Crosscap formula & Klein bottle & CP-group  & case \\ \hline
D & odd & \cite{FHSSW} & $ +++$ & $SO$ & 1 \\
  &    &   & $ ++- $ & $SO$ & 2 \\ \hline
C=T &  odd & Cardy/Rome & $+-0$ & $SO$ &  4  \\
    &     &      & $-+0 $   & $U$ & 5 \\ \hline
D=C &  even & Cardy/Rome  & $ +++$ & $SO$  &  1 \\
    &       &       & $++- $ & $SO$ & 2 \\
    &       &       & $-+0 $  & $U$ & 3 \\ \hline
T   &  even & exceptional & $ +-0 $ & $SO$ & 6 \\ \hline
\end{tabular} 
\end{center} 
\vskip .3truecm

Note that all allowed continuous Klein bottle amplitudes make their
appearance for both odd and even $N$, but in rather different ways.  Note
also that for even $N$ the diagonal invariant (D) allows four different
orientifold projections (the case $K_{-+0}$ actually consists of two
subcases with opposite signs for all Klein bottle coefficients in the
twisted sector), whereas the twist conjugation invariant (T) allows only one. This is
strange because we expect these theories to be dual to each other (in the
sense of the existence of a one-to-one map between their operators,
respecting all correlators).  This duality is of course not the T-duality
of the circle (which was modded out in the orbifold). We do not know if
such a duality has been proved in the literature, but it certainly seems to
hold in the simplest case, $N=2$, the tensor product of two Ising models. 
Note that T-dual circles admit the same number (namely two)
of O-plane/D-brane configurations, and
the only aspect that differs is the number of allowed D-brane positions. In the 
orbifold case two probably dual theories have a different number of orientifold
projections, corresponding to physically different
configurations, with different CP groups.
Although this is counterintuitive, on the other hand it does not seem to
contradict the duality in an obvious way. Note that also the number of boundary
conditions differs for the two mutually dual cases, but this merely
corresponds to a different choice of rationally allowed positions for the
same D-branes. Note furthermore that for T-dual rational circle theories the
number of orientifold choices is the same.

After this enumeration (which is exhaustive for small $N$)
only four of the twelve potential Klein bottles are realized. Most absences can be
explained by a combination of the following facts
 
\begin{itemize}
\item
The twist fields $\sigma_i, \tau_i$ and the fields
$\phi_i$ must be simultaneously (off)-diagonal in any MIPF.  This follows
from modular invariance. 

\item The Klein bottle coefficients of $\phi_1$ and $\phi_2$ must be identical,
   because these fields are either each others conjugates, 
   or they fuse to $\partial X \bar \partial X$,
   which must have projection sign 1. We call these coefficients $K_{\phi}$.

\item $K_{\phi}$ can be expressed as $\half(\epsilon_2+\epsilon_1^N)$.

\item The fusion of $\sigma_1$ and $\sigma_2$ produces fields $\varphi^k$,
   with $k+N$ odd.

\item The fusion of $\sigma_1$ and $\tau_1$ produces fields $\varphi^k$,
   with $k+N$ even.
\end{itemize}

The last two points are relevant only if the twist fields
appear diagonally. These points imply the following.
For conjugation invariants we must have $\epsilon_3=0$ and $\epsilon_2=-\epsilon_1^N$.
For diagonal invariants we must have $\epsilon_2=\epsilon_1^N$. Furthermore from point 5
we find that $\epsilon_1=1$ for $N$ odd. Hence $\epsilon_2=1$ for all diagonal invariants.
If $\epsilon_3=0$ the projections of
$\sigma_1$ and $\sigma_2$ are opposite. This is impossible for $N$ odd, and implies
$\epsilon_1=-1$ for $N$ even. On the other hand, if
$\epsilon_3 \not =0$, point 4 implies $\epsilon_1=\epsilon_2=1$
for $N$ even. 

This only leaves one case that was not found, and is also not yet ruled out,
namely $\epsilon_1=\epsilon_2=-1$ for conjugation invariants with even $N$.  This
case can be ruled out by computing the transverse channel amplitude. It turns out
that $\phi_1$ and $\phi_2$ propagate in the transverse channel, although they are
not Ishibashi states. Hence this case must be rejected.

As an additional check on the result one can now solve the U-NIMrep equations
for all other accessible MIPFs as well. The results are in complete agreement
with the foregoing: $Z_D(p,N)$ has 4 U-NIMreps for $N$ even, 2 for $N$ odd;
$Z_C(p,N)$ has one U-NIMrep for $N$ even, and 2 for $N$ odd; $Z_{ii}(p,N)$
always has 4 U-NIMreps, whereas $Z_{ij}(p,n), i\not=j$, has none. Finally
$Z_X(p,N)$ always has 4 NIMreps, except when $N=p^2$, in which case it has two.
The four U-NIMreps of $Z_X$ correspond precisely to two distinct Klein bottle
choices for the circle and its dual. For $N=p^2$ one ends up in the self-dual point,
which explains why one gets only half the number of solutions. In all cases these
distinct U-NIMreps correspond to choosing different boundary conjugations for
a single NIMrep, although $Z_C(3,9)$ and $Z_D(3,9)$ have respectively one
and two additional NIMreps that do not admit any U-NIMreps, and are presumably
spurious. Finally, $Z_{ij}(p,n), i\not=j$ was found to have a single NIMrep which
does not admit a U-NIMrep, in agreement with the fact that these MIPFs are
asymmetric.

The U-NIMreps for $Z_C(p,pq)$ and $Z_D(p,pq)$, with $p$ and $q$ prime have been
given explicitly in \cite{NunoSch}, and U-NIMreps for simple current MIPFs are
described in \cite{FHSSW}. In all other cases these conclusions, as well as
the completeness of the entire picture, rely on extrapolation to arbitrary $N$.

\subsubsection{Localization and Chan-Paton groups}\label{subsec:localization}

In the rational CFT description one can attempt
to get information about the boundary and crosscap states by analyzing the
Fourier transformation of their coupling to closed string states. 
This amounts to probing the brane/plane positions by scattering of gravitons \cite{Vecchia}
(or, equivalently, dilatons or tachyons).
This method has a clear physical interpretation in flat space, but
becomes less intuitive when applied to compact spaces, although sensible
results are obtained for the circle (as explained above) and WZW-models \cite{FFFS} \cite{HSS}.
Apart from the proper physical interpretation, a second caveat is that this method
requires a precise knowledge of the boundary and crosscap coefficients. In many cases
the latter are obtained by imposing integrality conditions on annulus, M\"obius and
Klein bottle amplitudes. These amplitudes are not sensitive to sign changes in the coupling
to Ishibashi states provided one makes the same sign change in the boundary and
the crosscap coefficients. Such sign changes do not affect tadpole cancellation either.
In principle the true sign can be determined by solving
the sewing constraints, but that has been done only in very few cases. However, 
in the Cardy case the results of \cite{FFFSA} imply that the signs are correct. This
should then also apply to all possible choices of orientifold projections, since this
is expected to add O-planes in different positions while keeping the branes fixed.

Keeping these caveats in mind, we can compute the positions as follows. The coupling to the
fields $\varphi^k$ provides a natural set of Fourier components for the couplings. Inspired
by the circle results we define a shape function
\begin{equation}
F(x)=\sum_{k=1}^{N-1} (e^{ikx/R} + e^{-ikx/R}) C(k) \ ,
\end{equation}
where $C(k)$ is a boundary or crosscap coefficient\footnote{We use here the
coefficients specified in \cite{FHSSW}, but without the denominator factors
$\sqrt{S_{Ki}}$. It is more natural to absorb these factors 
in the Ishibashi metric for the unoriented annulus, so that the boundary coefficients
themselves are independent of the choice of orientifold. See also \cite{FFFS}, \cite{HSS} and \cite{NunoSch}.}
for the coupling to $\varphi^k$.
This function is periodic with period $2\pi R$ and symmetric in $x \to -x$ as well
as $\pi R + x \to \pi R - x$, and therefore
it is natural to identify the line segment $[0,\pi R]$ with the orbifold. Note that in the diagonal
and conjugation modular invariants of the orbifolds with $R^2=\alpha' N$ all Ishibashi labels $k$
occur, so that there are no other points with reflection symmetry:
the identification of the two orbifold points is unambiguous.
By defining
$C(-k)=C(k)$ the second term can be used to extend the sum to negative $k$. The coefficients 
$C(0)-C(\partial X)$ and $C(\phi_1)+C(\phi_2)$ turn out to have precisely the right value
 to complete the sum to the range $-N \leq k < N$. The resulting function $F(x)$ typically
has one or two positive or negative peaks along the orbifold line, which approach $\delta$-functions
for large $N$. We interpret the extrema as $\rm{O}0$ plane positions, and the sign as an $\rm{O}0$ charge.
For the six cases in the table, the coefficients $C(k)$
for the crosscaps either vanish for all even $k$, or
for all odd $k$. The non-vanishing values are all equal to ${1/\sqrt{N}}$, up to a sign. If this
sign is positive,
this leads respectively to opposite charge or same-charge planes at the two
endpoints of the orbifold line, $x=0$ and $x=\pi R$.
The other possibility we encounter is a sign $(-1)^{k/2}$ for even $k$.
This shifts the plane positions by $\half \pi R$, so that they are on top of each other.

Because the orbifold incorporates the circle T-duality, which interchanges $\rm{D}0 (\rm{O}0)$ with
$\rm{D}1 (\rm{O}1)$ branes (planes) we expect boundary and crosscap states to describe a combination
of branes and planes of dimension 0 and 1. While the $\rm{D}0/\rm{O}0$ positions and charges can be extracted
very easily from $F(x)$, this is not the best way to determine the $\rm{D}1/\rm{O}1$ charges.  The information
is in fact hidden
in the linear combinations $C(0)+C(\partial X)$ and $C(\phi_1)-C(\phi_2)$ which are not used in the
computation of $F(x)$. A Fourier transform of these two quantities yields identical values
on all allowed brane positions in the first case, and alternating values on even and odd positions in the second case. 
Furthermore in all cases either $C(0)+C(\partial X)$ or $C(\phi_1)-C(\phi_2)$ is zero. Remembering
how $\rm{D}1$ branes emerged for the rational circle, we are led to the conclusion that a non-vanishing
$C(0)+C(\partial X)$ implies the presence of two equal-charge $\rm{D}1/\rm{O}1$ branes/planes, whereas a non-vanishing
value of $C(\phi_1)-C(\phi_2)$ implies two opposite-charge $\rm{D}1/\rm{O}1$ branes/planes.

The ``charge" referred to above always refers back to the corresponding quantity for the
circle, namely the dilaton coupling strength. This allows us to interpret any
orbifold brane/plane configuration
in terms of a collection of circle configurations of different dimension and charges.
Not surprisingly, this interpretation breaks down for branes labelled by twist fields, that
have no circle analog.
Furthermore the values $C(\sigma_i)$ and $C(\tau_i)$ (which vanish for crosscaps and most
boundaries) are also not used, and we do not have a geometric interpretation for these values.

Some more information about brane and plane positions can be gathered from boundary conjugation,
which geometrically corresponds to a reflection of a brane through an O-plane. This property
affects
the Chan-Paton groups of the brane, which is orthogonal or symplectic for self-conjugate (real) branes
and unitary for pairs of conjugate branes. Boundary conjugation and the allowed CP-groups
are not affected by the aforementioned sign ambiguities, but the distinction between symplectic
and orthogonal for real boundaries does depend on the overall sign of all crosscap coefficients
relative to all boundary coefficients.
This sign determines the O-plane tension,
and whenever we specify a CP group below
we have fixed the tension to a negative value, so that the dilaton tadpole can be cancelled (in principle)
between D-branes and the O-plane. Orbifold O-planes (unlike circle O-planes) always have non-zero tension.

For the cases discussed listed in the table we find the following
positions:
 
\begin{itemize}
\item
N even, Diagonal invariant: This is the Cardy case, so boundary labels correspond
to primary labels, and the localization analysis should be reliable. The branes with 
labels $0$ and $\partial X$ are at $x=0$, the ones with label $\phi_i$ are at $x=\pi R$.
The branes with labels $\varphi^k$ are localized at points equally spread over the interval.
All these branes have in addition a D1 component. The circle-inspired Fourier analysis 
cannot be trusted for the twisted sector branes and indeed gives contradictory results.
The orientifold choices
correspond to the four distinct choices of  the Klein bottle current, $K=0,\partial X, \phi_1$
and $\phi_2$. For $K=0$ we get $K_{+++}$, and we find two $\rm{O}0_{+}$-planes at the orbifold points plus
two $\rm{O}1_{+}$-planes; For $K=\partial X$ ($K_{++-}$) we get two $\rm{O}0_{-}$ planes at the orbifold points, and
again two $\rm{O}1_{+}$-planes. For $K=\phi_1$ or $\phi_2$ ($K_{-+0}$) we get two coincident $\rm{O}0_{+}$-planes
at $x=\half \pi R$, plus an $(\rm{O}1_{+}+\rm{O}1_{-})$ configuration.
For $K_{+++}$
all CP-groups are $SO$. For $K_{++-}$
the CP groups of boundaries $0,\partial X, \phi_i, \sigma_i$ and $\tau_i$ become unitary, while
all others remain $SO$. For $K_{-+0}$ 
all CP groups are unitary, except for $\varphi^k, k=N/2$ and the twist fields
with either label 1 or 2, for which we find orthogonal groups. The group type for the $\varphi^{k}$
branes is easy to understand: if the O-planes are in the middle of the orbifold line segment, they
conjugate the branes mutually, except the brane in the center, which is self-conjugate. If the planes
are on the endpoints, they conjugate all $\varphi$-branes to their orbifold image, {\it i.e.} to
themselves, so that they are self-conjugate. A clear geometric picture suggests itself. Given a
choice for the orbifold plane, there are two choices for the orientifold plane: on top of it,
or orthogonal to it. The first choice leads to $K_{+++}$ and $K_{++-}$ and
mostly self-conjugate branes, the second to $K_{-+0}$ and mostly conjugate brane pairs.  The 
proper geometric interpretation of the CP groups of the
eight ``special" branes (those not labelled by $\varphi^k$) is somewhat less
intuitive.

\item
N even, Twist automorphism: This is an exceptional invariant, and we obtained the
boundary and crosscap coefficients numerically for small values of $N$, up to the
sign ambiguity described above. Given the Klein bottle amplitude
 ($K_{+-0}$) one can easily compute the crosscap coefficients for all $N$:
 $ C(0)=C(\partial \phi)=\half,\ \ C(\varphi^{2k+1}) = {1\over \sqrt{N}}, C(\varphi^{2k}) = 0$
This implies an $\rm{O}0_{+}$ plane at $x=0$ and and $\rm{O}0_{-}$ plane at $x=\pi R$. 
In addition there are two $\rm{O}1_{+}$ planes.
All CP groups are
orthogonal. We have no explicit formula for
the reflection coefficient for arbitrary $N$, although it could be obtained in principle using the methods of
\cite{Birke} applied to the twist orbifold of the $c=1$ orbifold (which is the $c=1$ orbifold with
four times the value of $N$). In the absence of such a formula it is difficult to discuss brane
positions with these methods. There is also no canonical labelling of the boundary states.

\item
N odd, Charge conjugation invariant. The discussion of brane positions is identical
to the one for even $N$, except that there is no brane in the middle. There are four
possible choices for the Klein bottle current, but $K=\partial X$ and $K=\phi_2$ are
known to be identical to $K=0, K=\phi_1$ respectively, up to interchange of branes  \cite{FHSSW}.
For $K=0$ ($K_{+-0}$) the O-plane configuration is as for even $N$, and
all CP groups are orthogonal except those of the $\phi_i, \sigma_i$ and $\tau_i$ branes, which
are unitary.  For $K=1$ ($K_{-+0}$) the O-plane configuration is also the same as for even $N$, and all CP-groups
are unitary except the ones labelled by $\sigma_i,\tau_i$ ($i=1$ or 2), which are orthogonal. Apart from the usual
eight special branes, these results are analogous to those for even $N$.

\item
N odd, Diagonal invariant. Here the formulas of \cite{FHSSW} apply.
The boundary states are labelled by orbits of the simple current $\phi_1$.
There are $N+1$ branes,
two for each label $\varphi^k$ with k even, one labelled by $0$, and one labelled
by a twist field. Boundary ``0" is localized at the orbifold endpoints, the twist
field boundary is not localizable, and all other boundaries occupy two symmetric
positions on both sides of the center. There are two orientifold projections,
distinguished by opposite crosscap signs. One of them yields $K_{+++}$, and
all CP-groups are $SO$. The other yields $K_{++-}$, and all groups are $SO$ except
the one of the twist field boundary, which is symplectic. 
The O-plane configuration consists of two $\rm{O}1_{+}$ planes, plus two $\rm{O}0_{+}$ (for $K_{+++}$)
and two $\rm{O}0_{-}$ (for $K_{++-}$) planes located at the center.
The fact that the boundaries
are self-conjugate is understood as a consequence of the fact that each is
symmetrically located on each side of the $\rm{O}0$-plane. Note however that the picture
seems rather different than for even $N$, where the $\rm{O}0$ planes are at the orbifold points.
Note also that in this case the caveat
regarding signs of the coefficient applies. If we modify all coefficients $C(k)$
by a factor $(-)^{k/2}$ the O-plane positions are as for even $N$ (however, the brane
positions, which also change, are still different than they are for even $N$).
\end{itemize}

Finally we can extract from \cite{NunoSch} the crosscap and boundary coefficients
for $R^2=\alpha' p/q$, $pq$ odd, but only up to signs, as explained above.
For the crosscaps the Fourier transformations splits naturally into two sums,
one proportional to $1/\sqrt{q}$ and the other to $1/\sqrt{p}$. The first
gives $\rm{O}0$ planes at multiples of $1/q$ of the full radius, the second
at multiples of $1/p$, with signs depending on the case considered.
These sums are completed by including $C(0)-C(\partial X)$,  $C(\phi_1)+C(\phi_2)$
in one of the sums and $C(0)+C(\partial X)$, $C(\phi_1)-C(\phi_2)$ in the other, in
agreement with the foregoing discussion.
The result
can be interpreted either in terms  of a circle of radius $R^2=\alpha' p/q$  or in terms
of a circle  of radius $R^2=\alpha' q/p$. The first possibility corresponds a $q$-fold identification
of the orbifold line, the second to a $p$-fold identification. In the first case the planes
originating from the first Fourier sum are at the endpoints, whereas those from the second
one are distributed equally on $p$ points of the reduced line segment. It is natural to
regard the latter as rational CFT realizations of $\rm{D}1$ branes. 

For $Z_C(p,q)$ two orientifold
projections were found in \cite{NunoSch}, that differ by interchanging $p$ and $q$. The O-plane
charges are alternating for one of the Fourier sums, and identical for the other. On the
reduced orbifold line segment this can be interpreted as a configuration with two $\rm{O}0_{+}$ planes
at the end, plus one $\rm{O}1_{+}$ and one $\rm{O}1_{-}$ plane, and a configuration with one $\rm{O}0_{+}$ plane
and one $\rm{O}0_{-}$ plane at the end, plus two $\rm{O}1_{+}$ planes (two, because odd and even points are to be
identified with different $\rm{O}1$-planes, as in the previous case).
For $Z_D(p,q)$ there are also two orientifold projections, this time differing by signs in 
the crosscap coefficients, that flip the two Fourier sums with
respect to each other. Using the same interpretation we now get two $\rm{O}0_{+}$ planes at the end, combined
with
either two $\rm{O}1_{+}$ or two $\rm{O}1_{-}$ planes. All this is identical to
the results for odd, integer
radius, except for the positions of the two $\rm{O}0_{+}$ planes. 
But precisely these positions are affected by
the unknown signs. This particular kind of simple current MIPF 
(generated by a ${\Zbf_4}$-current with fixed points)
does not appear
in the circle theory and hence the correctness of these signs cannot be tested using brane
localization on the circle.

\subsection{Orientifolds for arbitrary radius}\label{sec:geometrical-approach}
In this subsection we argue that the four Klein bottle amplitudes that we obtained in
the previous subsection are the only possible ones for arbitrary radius.
Thus, we find all possible orientifold
maps, that is maps that project out states that are not invariant under
the exchange $z \leftrightarrow \zb$.
Since the orientifold transformation of $X_L$ and $X_R$ must square to the identity, 
the oscillators transform like $a_{n} \to \ab_n$.  The only
freedom left is in the operators coming from the $z$ independent
parts in the expansion of $X_L$ and $X_R$.

The standard orientifold projection \cite{ps}
corresponds to the map $X_L \to X_R$, giving rise to the amplitude
(see \eqref{klein_orb})
\begin{equation}
K_{+++}=
\half \frac{1}{\eta(2i\tau_2)} \sum_{k\in\Zbf} q^{\frac{\alphap}{2}\left(\frac{k}{R}\right)^2}
+ \half \frac{1}{\eta(2i\tau_2)} \sum_{m\in\Zbf}
q^{\frac{1}{2\alphap}\left(mR\right)^2}+ 
2 \sqrt{ \frac{\eta}{\theta_4} } \ .
\end{equation}
The first two terms arise from a trace over the states in the untwisted
sectors. In the first of these two the orientifold map is inserted, and only the
states with zero winding contribute. The second is the one with both the
orientifold and orbifold map inserted, and since the KK momentum is not invariant
under reflections, only states with no KK momentum contribute. The last term
comes from the two twisted sectors; they both contribute in the same
amount.

The first variation is to let the operators that create
the ground states in the twisted sectors acquire a minus sign under the
orientifold transformation. This will result in a minus sign in the last term
of \eqref{klein_orb}, giving
\begin{equation}
K_{++-}=
\half \frac{1}{\eta(2i\tau_2)} \sum_{k\in\Zbf} q^{\frac{\alphap}{2}\left(\frac{k}{R}\right)^2}
+ \half \frac{1}{\eta(2i\tau_2)} \sum_{m\in\Zbf}
q^{\frac{1}{2\alphap}\left(mR\right)^2} -
2 \sqrt{ \frac{\eta}{\theta_4} } \ .
\end{equation}

In order to understand which other possible maps are allowed, one has to 
consider the way the various sectors interact. The untwisted sectors combine
according to
\begin{equation}
(+,n)(+,m)=(+,n+m) \ . \label{eq:wind_add}
\end{equation}
Apart from the standard projection, this equation allows for the map
\begin{equation}
\begin{split}
X_L \to X_R + \frac{\pi}{2} \frac{\alphap}{R} \\
X_R \to X_L - \frac{\pi}{2} \frac{\alphap}{R} \ ,
\end{split}
\label{windingshift}
\end{equation}
that changes the sign of the states in the sectors with odd winding.
Because twisted sectors combine like
\begin{equation} \label{eq:-oddx-odd=+even}
(-,\hbox{odd})(-,\hbox{even})=(+,\hbox{odd})\ ,
\end{equation}
consistency requires that the two twisted sectors contribute with
opposite sign after the projection. The resulting amplitude is 
\begin{equation}
K_{+-0}=
\half \frac{1}{\eta(2i\tau_2)} \sum_{k\in\Zbf} q^{\frac{\alphap}{2}\left(\frac{k}{R}\right)^2}
+ \half \frac{1}{\eta(2i\tau_2)} \sum_{m\in\Zbf} \left( -1 \right)^m
q^{\frac{1}{2\alphap}\left(mR\right)^2} 
\ ,
\end{equation}
where the contribution from the twisted sectors cancels. 

>From T-duality one can then obtain the last consistent map,
\begin{equation}\begin{split}
X_L \to X_R + \frac{\pi}{2} R \\
X_R \to X_L + \frac{\pi}{2} R \ ,
\end{split}
\label{momentumshift}
\end{equation}
that changes the sign of states with odd KK momentum. Since $X \to X +
\pi R$, the twisted ground state localized in $X=0$ is swapped for the
one localized in $X=\pi R$. This means that 0 and $\pi R$ are no longer fixed points
of the orientifold map, whose eigenstates are now localized in $X=\pm \frac{\pi}{2} R$.
Moreover, since the trace over the twisted states vanishes after the projection, there
is no contribution from the twisted
sectors to the Klein bottle amplitude, whose form is 
\begin{equation}
K_{-+0}=
\half \frac{1}{\eta(2i\tau_2)} \sum_{k\in\Zbf} \left( -1 \right)^k q^{\frac{\alphap}{2}\left(\frac{k}{R}\right)^2}
+ \half \frac{1}{\eta(2i\tau_2)} \sum_{m\in\Zbf} 
q^{\frac{1}{2\alphap}\left(mR\right)^2} 
\ .
\end{equation}

In order to understand why the choice $\epsilon_1 = \epsilon_2 = -1$ in eq. \eqref{klein_orb} is not
allowed, we have to analyze the transverse channel.  We only need to
consider the untwisted sector, so we concentrate on the case $K_{--0}$.  
The Klein bottle in the
direct channel depends on $2 i \tau_2$, the modulus of the
doubly-covering torus. The Klein bottle in the transverse channel is
obtained performing an $S$ modular transformation on the modulus of
the doubly-covering torus, that is writing the amplitude in terms of $
\ell = 1/ 2\tau_2$, the proper time in the transverse channel,
describing a closed string propagating between two orientifold planes.
The end-result of this transformation (see for instance
\cite{AngelSagn} for a review) is
\begin{equation}
\begin{split}
&\tilde{K}_{++\pm} =
{{\coeff12} } {1\over \eta(i\ell)} R \sqrt{2 \over \alphap} 
\sum_{k\in\Zbf} q^{{1 \over \alphap}(k R)^2}
  + {\coeff12}{1\over \eta(i \ell)} {1 \over R} \sqrt{2 \alphap}
\sum_{m\in\Zbf} q^{{\alpha'}({m \over R})^2} \pm 2 \sqrt{\frac{\eta}{\theta_2}}
\\ 
&\tilde{K}_{+-0} =
{{\coeff12} } {1\over \eta(i\ell)} R \sqrt{2 \over \alphap} 
\sum_{k\in\Zbf} q^{{1 \over \alphap}(k R)^2}
  + {\coeff12}{1\over \eta(i \ell)} {1 \over R} \sqrt{2 \alphap}
\sum_{m\in\Zbf} q^{{\alpha'}({{m +1/2} \over R})^2}
\\
&\tilde{K}_{-+0} =
{{\coeff12} } {1\over \eta(i\ell)} R \sqrt{2 \over \alphap} 
\sum_{k\in\Zbf} q^{{1 \over \alphap}(k+ 1/2)^2 R^2}
  + {\coeff12}{1\over \eta(i \ell)} {1 \over R} \sqrt{2 \alphap}
\sum_{m\in\Zbf} q^{{\alpha'}({{m} \over R})^2}
\\ 
&\tilde{K}_{--0}=
{{\coeff12} } {1\over \eta(i\ell)} R \sqrt{2 \over \alphap} 
\sum_{k\in\Zbf} q^{{1 \over \alphap}(k+ 1/2)^2 R^2}
  + {\coeff12}{1\over \eta(i \ell)} {1 \over R} \sqrt{2 \alphap}
\sum_{m\in\Zbf} q^{{\alpha'}({{m+1/2} \over R})^2} \ , 
\end{split}
\end{equation}
where here $q = e^{-2 \pi \ell}$. The states that contribute to 
the Klein bottle amplitude in the transverse channel are 
closed-string states propagating between two orientifold planes. 
In $\tilde{K}_{++\pm}$ only states with even KK momentum or even 
winding contribute in the untwisted sector. In $\tilde{K}_{+-0}$ only states with odd KK momentum
or even winding contribute. This is consistent with the direct channel,
since $K_{+-0}$ projects out only states with odd winding.
The same is valid for $\tilde{K}_{-+0}$, where only states with
even KK momentum and odd winding contribute, while in the direct 
channel the states with odd KK momentum are projected out.
Finally, in the case of $ \tilde{K}_{--0}$ states with odd KK momentum
and states with odd winding contribute to the transverse amplitude, 
but these states are both projected out by $K_{--0}$, so that 
this Klein bottle projection is not consistent.

>From the transverse channel analysis one can also derive the position 
of the orientifold planes for the various Klein bottles.
In all cases, namely the standard orientifold projection $X_L \to X_R$, and the ones
given in \eqref{windingshift} and \eqref{momentumshift}, the map has no fixed points,
and this corresponds to introducing O1-planes. Once the orbifold map $X\to -X$ is implemented,
all these maps develop fixed points, where O0-planes are located.
This means that all these amplitudes describe a configuration of two O1 planes, and two O0 planes 
located at the fixed points of the `orientifold+orbifold' ($\Omega r$) map. 
The twisted sector corresponds in the transverse 
channel to closed string states propagating between an O1 plane and an O0 plane.

In the case of $K_{++\pm}$, the $\Omega$ and $\Omega r$ maps are respectively
$X \to X$ and $X\to -X$, and
the two O1 planes have the same tension, as well as
the two O0 planes. The two sign options for the twisted sector correspond to the fact that 
the tension of the O0 planes can be positive or negative with respect to the 
tension of the O1 planes\footnote{The $K_{++-}$ case is analogous to the six-dimensional 
brane supersymmetry breaking model of \cite{ads}, in which the O5-planes and the O9-planes 
have opposite tension.}. 
Since the $\Omega r$  map has fixed points in 0 and $\pi R$, the 
position of the two O0-planes coincides with the two fixed points of the orbifold.
Applying the T-duality transformation $X = X_L + X_R \to X' =X_L - X_R$, and $R \to R'=\alphap / R$, one can see
that these two Klein bottles are both self-T-dual.
In the case of $K_{+-0}$, the $\Omega$ projection is given in \eqref{windingshift}, and it maps $X$ to itself,
meaning that the two O1 planes have the same tension.
The $\Omega r$ projection maps $X$ to $-X$, so that the O0-planes are located at the fixed points 
of the orbifold. In order to determine the tension of the O0-planes, one has to consider how the 
$\Omega r$ transformation acts on the T-dual coordinate. In this case one has $X' \to X' +\pi R'$, 
and the shift in the dual coordinate is a manifestation of the fact that the two O0-planes have 
opposite tension. The Klein bottle amplitude corresponds thus to the configuration $\rm{O}1_+ \oplus \rm{O}1_+ \oplus \rm{O}0_+
\oplus \rm{O}0_-$, and
the twisted sector cancels because of the opposite contribution
from the two orbifold fixed points, where the O0-planes are located.
Finally, in the case of $K_{-+0}$, the $\Omega$ projection is given in \eqref{momentumshift}, and it maps $X$ 
to $X + \pi R$, so that the two O1-planes have opposite tension, while $\Omega r$ maps $X$ to $-X =\pi R$,
so that the O0-planes are located in the middle of the orbifold segment.
In this case the twisted sector  cancels separately in any of the two (coincident)
orientifold fixed points. 
T-duality maps this configuration to the previous one. 

In all these cases the locations and charges of the O0 planes and the charges of the O1 planes agree
with the results obtained from the CFT analysis in subsection \ref{subsec:localization}, except those
for the $K_{++\pm}$ Klein bottle of the D-invariants for odd $N$. In that case the charges are the same,
but the two $\rm{O}0_{+}$ planes were found in the center rather than at the edges. But this was precisely
a non-Cardy case, where the signs of the crosscap coefficients (crucial for the precise location) are
not determined. 

The foregoing discussion was for arbitrary radius, and seemed to rely in all cases on the
standard orbifold map $X\to -X$. At rational radii the various orientifolds occur in combinations
with specific partition functions, which require different orbifold maps, discussed in the appendix.
This changes the map $r$ in the foregoing discussion, and hence its fixed points. However, it
also changes $\Omega r$, whose fixed points determine the O-planes. It is easy to see that both
modifications cancel, so that the relative position of orbifold fixed points and $\rm{O}0$-plane 
positions remains unchanged. Note that nothing in the analysis in this section imposed any
relation between the orientifold map and the orbifold map in rational points. This relation was found
in subsection \ref{subsec:unimreps}\ and makes use of OPE's involving distinct twist fields.
We did not consider twist fields in this section, and furthermore for non-rational radius they
cannot be distinguished, hence there is no reason to expect such a relation to emerge.

\subsection{Orientifolds of exceptional MIPFs}\label{sec:orient-except-mipf} 

As mentioned in section \ref{sec:cft-approach}, the orbifold has 
exceptional MIPFs, constructed using an
automorphism, $\omega$, which leaves the fusion
coefficients invariant:
$N_{ij}^{\phantom{ij}k}=N_{\omega(i)\omega(j)}^{\phantom{\omega(i)\omega(j)}\omega(k)}$.
The exceptional torus partition functions obtained from the chiral algebra
of the CFT at square radius $R^2 = \alphap p q$, with $p$ and $q$ odd prime numbers, are
\begin{equation}
T=\sum\chi_{i}\delta_{i\omega(j)}\bar\chi_j \ , \quad 
T=\sum\chi_{i} C_{i\omega(j)}\bar\chi_j \ .
\end{equation}
In \cite{NunoSch}, these two invariants were called ``diagonal + automorphism'' (D+A)
and ``Cardy + automorphism'' (C+A) respectively. Geometrically, these two tori describe 
a free boson compactified on an orbifold of radius $R^2 = \alphap p/q$ and its T-dual.

We first review the results of \cite{NunoSch} about orientifold projections.
In the D+A case, the trivial Klein bottle, that is 
$K_i =1$ for all the fields that couple diagonally on the torus, is allowed. A
second Klein bottle is also allowed, with $K_i = -1$ for the twist fields
and $K_i =1$ for the other diagonal fields.
In the C+A case, surprisingly the trivial choice $K_i=1$ for all the 
fields coupling diagonally on the torus is not allowed. There 
are two Klein bottles. One has $K_{\phi_k} = -1$ when $k$ is an odd multiple of $p$ 
and $K_i =1 $ otherwise, and the other is obtained exchanging $p$ with $q$.

Looking at the resulting amplitudes as functions of the orbifold radius, 
one realized that these Klein bottles are precisely the ones obtained in section 
\ref{sec:geometrical-approach}, for a bosonic string compactified on a circle of 
square radius $R^2 = \alphap p/q$. 
In particular, the Klein bottles of the 
D+A modular invariant are $K_{+++}$ and $K_{++-}$, while the
Klein bottles of the C+A modular invariant are $K_{+-0}$ and $K_{-+0}$. Since the twisted 
sector is not diagonal for the C modular invariant, this is the only possibility that is allowed in light
of the results of the previous subsection, and thus the results of \cite{NunoSch} are
completely consistent with the orientifold projections that are allowed for arbitrary
radius.

\section{Conclusions}
 
We have identified four distinct orientifold projections for the $c=1$ orbifolds.
Geometrically, they can be described most easily starting from the $\rm{O}0$ plane
configurations of the T-dual circle. The $(\rm{O}0_{+},\rm{O}0_{-})$ configuration has only
one axis of symmetry, namely the line through the $\rm{O}0$-planes. Hence the orbifold
and O-plane directions must line up, and the orbifold ${\rm{O}}0$ planes are at its endpoints.
The configuration $(\rm{O}0_{+},\rm{O}0_{+})$ has two axes of symmetry, and the orbifold
reflection line is either on top of or orthogonal to the orientifold line. Then the O0-planes
are respectively at the endpoints or on top of each other in the center.

In the circle theory on can distinguish two T-dual orientifold maps, one of the
form $X_L \to +X_R + {\rm const}$ and one of the form $X_L \to -X_R + {\rm const}$.
The former has
fixed points in $X_L-X_R$, but not in $X_L+X_R$, whereas for the latter it is just
the other way around. Therefore the former gives rise to ${\rm O}1$ planes on the circle and the latter
to ${\rm O}0$-planes on the T-dual circle.
The orbifold map (which has fixed points both in $X_L-X_R$ and $X_L+X_R$)
transforms the two types of orientifold maps into each other, so that both ${\rm O}1$ and ${\rm O}0$
planes are present. Inspection of the transverse channel show that
the charges of the O1 planes are identical if the orbifold fixed plane and
the orientifold plane coincide, whereas they are opposite if these fixed planes are orthogonal.
Allowing for an additional relative sign between the ${\rm O}1$ and ${\rm O}0$ planes then gives
a total of four configurations (since the overall sign is irrelevant):
($\rm{O}1_+ \oplus \rm{O}1_+) \oplus (\rm{O}0_+\oplus \rm{O}0_+$),
($\rm{O}1_+ \oplus \rm{O}1_+) \oplus (\rm{O}0_-\oplus \rm{O}0_-$),
($\rm{O}1_+ \oplus \rm{O}1_+) \oplus (\rm{O}0_+\oplus \rm{O}0_-$) and
($\rm{O}1_+ \oplus \rm{O}1_-) \oplus (\rm{O}0_+\oplus \rm{O}0_+$).  This argument also shows
why a fifth logical possibility, ($\rm{O}1_+ \oplus \rm{O}1_-) \oplus (\rm{O}0_+\oplus \rm{O}0_-$),
cannot occur.

This intuitive
argument was worked out in detail in section \ref{sec:geometrical-approach}, and is
backed up by the complete solution for U-NIMreps for rational CFT.
The latter classification can be done exhaustively, but this is
necessarily limited to a few rational points. We have shown that all four orientifolds
are realized in all rational points, although in rather different ways. We have also
shown how a known, but initially surprising solution at $R^2=\alpha' p/q$ fits in
perfectly with the continuum.  

At arbitrary $R$
we cannot rigorously rule out additional solutions,
but in view of the agreement between the continuous $R$ and the rational CFT
descriptions, any deviations would be quite surprising.  

A few open problems remain. While all methods agree on the O-plane charges, there is
a discrepancy on their precise positions in one case, interestingly precisely the case were the CFT
results are least reliable. Secondly, the nature of the duality between diagonal and conjugation
invariants of the rational orbifolds needs to be clarified. Finally, in the geometric
description, applied to rational radii,
the link between the choice among those two invariants and the orientifold map
is not manifest.

\section*{Acknowledgments}\label{sec:acknowledgments}
We would like to thank Christoph Schweigert and Yassen
Stanev for discussions and Herman Verlinde for clarifying
some points in \cite{DVVV}. A.N.S. wishes to thank IMAFF-CSIC,
Madrid, where part of this work was done, for hospitality. The 
work of F.R. and A.N.S. has been performed as part of the program
FP 52 of the Foundation for Fundamental Research of Matter (FOM), and
the work of T.P.T.D. and A.N.S. has been performed as part of the 
program FP 57 of FOM.
The work of B.G.-R. and A.N.S. has been partially
supported by funding of the Spanish ``Ministerio de
Ciencia y Tecnolog\'\i a", Project BFM2002-03610.
\newpage

\section*{Appendix: Orbifold maps}
 \vskip .7truecm
Here we will discuss how the various partition functions enumerated in section \ref{Emi}
can be obtained from the circle theory. The standard description of orbifolds starts
with a circle theory, from which the $Z_2$-symmetry $X \to -X$ is modded out.
It is not hard to see that applying this map to the circle theories $Z(p,N)$ or
$Z(N/p,N)$ (with $1 \leq p \leq \sqrt{N}$ one obtains in both cases the orbifold theory
$Z_{\ldots}(p,N)$. The problem is that in the rational case
the orbifold partition function has an
additional label D, C or $ij$. Since the distinction is not made by the T-duality of the
circle, there must be more than one way to do the orbifold map. Obviously one can
generalize it to $X \to a-X$, {\it i.e.} rotating the plane of reflection, but this does
not have the desired effect.

It turns out that one must consider the chiral orbifold map
$$ X_L \to a_L-X_L \ ; \ \ \ \ X_R \to a_R-X_R\ . $$
On the vertex operators $V(k)$ corresponding to the fields $\varphi^k$ the only effect
is a phase between the two terms of which they consist; but the effect is more
important for the generators of the chiral algebra and the fields $\phi_1$ and $\phi_2$,
which make the difference between the rational and the non-rational case (we will ignore
the twist fields here, since the difference between the various partition functions is
already clear in the untwisted sector). Note that the circle theory operators from which
 $\phi_1$ and $\phi_2$ originate, which have $k=\pm N$ and chiral ground state multiplicity 2,
appear in an identical way for a circle and its T-dual.

The dependence of these vertex operators on $a_L$ and $a_R$ is as follows
for the chiral algebra generators
$$ W_L = V(R,0) + e^{i a_L{2R\over\alpha'}} V(-R,0)  $$
and
$$ W_R = V(0,R) + e^{i a_R {2R\over\alpha'}} V(0,-R) \ , $$
where
$$ V(r,s)=e^{i{2r\over \alpha'}X_L} e^{i{2s\over \alpha'}X_R}\ . $$
For the other four circle operators with $\mid r\mid =\mid s\mid=\half R$
we get two invariant combinations
 $$ V^A = V(+,+) +  e^{ i {R\over \alpha'}(a_L+a_R)} V(-,-)
 $$
and
  $$ V^B = V(+,-) +  e^{ i   {R\over \alpha'}(a_L-a_R)} V(-,+) \ , 
 $$
with the arguments ``$+$" and ``$-$" denoting $+R/2$ and $-R/2$ respectively.

The operators $V^A$ and $V^B$ are Virasoro-degenerate, but are distinguished by
the chiral algebra operators $W_L$ and $W_R$. In order to relate these operators to
a partition function interpretation we need to combine $V^A$ and $V^B$ into chiral
algebra eigenstates. For this we need the OPE of these operators, and here an important
r\^ole is played by the cocycle factors that should be added to these operators \cite{IKKS}.
In this case these cocycles can be conveniently represented by Pauli matrices
$(\sigma_3)^m (\sigma_1)^n$ where $m$ and $n$ are the winding and momentum quantum
numbers of the operator. It is easy to see that $W_L$ and $W_R$ acquire a factor 
$\sigma_3 (\sigma_1)^N$, $V(+,+)$ and $V(-,-)$ a factor $(\sigma_1)^N$ and
$V(+,-)$ and $V(-,+)$ a factor $\sigma_3$. Hence for even $N$ the cocycles do not change
anything in comparison with the ``naive" OPE.  For arbitrary $N$ the chiral algebra
eigenstates are found to be
 $$ V(+,+)+e^{i{R\over \alpha'}(a_L+a_R)}  V(-,-)\pm ( i^N e^{i {R\over \alpha'} a_R} V(+,-)
+ i^N e^{i {R\over \alpha'} a_L} V(-,+)) \ . $$
For even $N$ this can be factorized as
$$ \left[V_L(+)  \pm e^{i{R\over \alpha'}a_L} V_L(-)\right]\left[V_R(+)  \pm e^{i{R\over \alpha'}a_R} V_R(-)\right] \ ,$$
with correlated signs in the two factors;
for odd $N$ it cannot be factorized.
For these operators to have sensible reality properties, $a_L$ and $a_R$ must be quantized as
multiples of $\alpha' \pi/R$, the allowed positions in the
rational CFT description (these are precisely the allowed brane positions on
the circle; any other value would not allow a rational CFT interpretation). 
Then one finds
that for $N$ even the operators are real, and for $N$ odd they are each others conjugate, in agreement
with the modular matrix $S$ \cite{DVVV}.

For the standard case $a_L=a_R=0$, and for $N$ even, the operators have the expected ``cos cos"
and ``sin sin" form that is indicative of the diagonal invariant. By choosing $a_L=0$, $a_R=\alpha' \pi/R$
one can change this to a ``cos sin" and ``sin cos" form, corresponding to the conjugation invariant.
For odd $N$ the results are similar. The operators for $a_L=a_R=0$ can be written in the form
``cos cos $\pm i$ sin sin" and they change to ``cos sin $\pm i$ sin cos" for $a_L=0,  a_R=\alpha' \pi/R$.
These two cases should correspond, respectively, to the diagonal and charge conjugation invariant
of the odd $N$ orbifold. To get the heterotic orbifold invariants we may choose
$a_L=0, a_R= \alpha' \pi/2 R$. 
Note that this value for $a_R$ does not belong to the set of allowed positions, but it is an
allowed position for the orbifold with twice the value of $R$. The heterotic theory is obtained as
a chiral algebra extension of the latter CFT. The term in the partition function corresponding to
$V^A$ and $V^B$ has multiplicity 2, and it is a simple current fixed point, which
cannot be resolved using the orbifold data alone. Hence the reality properties of these operators
are not determined.

The cocycle factors are also needed in the operators $\Omega$ that implement the various orientifold
maps on the vertex operators. In some cases one has to include a factor $\sigma_3$ in these operators,
which affects the result only for odd $N$ and only when acting the operators $V^A$ and $V^B$.

\end{document}